\documentclass[twocolumn,amsmath,amssymb]{revtex4}
\usepackage{graphicx}
\begin{document}

\title{Quantum Entanglement in Heisenberg Antiferromagnets}
\author{ V. Subrahmanyam}
\address{Department of Physics, Indian Institute of Technology, Kanpur,
India.}
%\date{\today}
\begin{abstract} 
Entanglement sharing among pairs of spins in
Heisenberg antiferromagnets is investigated using the concurrence
measure. For a nondegenerate S=0 ground state, a simple formula relates
the concurrence to the diagonal correlation function. 
The concurrence length is seen to be extremely short.
Various finite clusters are studied numerically, to see
the trend in two dimensions. It is argued that the concurrences vanish
between pairs of spins which are not nearest neighbors, for the linear
chain and square lattice antiferromagnet ground states. For the triangular
lattice and Kagome' lattice, nearest-neighbor concurrences also vanish.
The concurrences in the maximal-spin states are explicitly calculated,
where the concurrence averaged over all pairs is larger than the S=0 states.
\end{abstract}
\maketitle
%\begin{multicols}{2}
\section{Introduction}

In recent years quantum entanglement has emerged as a common platform 
for scientists working in various fields such 
computer science,
physics, mathematics and chemistry\cite{Nielsen}.
In particular, quantum entanglement of spin-${1\over2}$ degrees of freedom, 
qubits, has
been studied extensively, due to their importance
for quantum computers, not to mention their well-known
applicability in various condensed-matter systems, optics and
other branches of physics. For a pure state of many qubits,
quantum entanglement, which is quantified by the von Neumann entropy of the
reduced density matrix, is a measure of how a subsystem is correlated
to the rest of the system. The key ingredient for entanglement is 
the superposition of states. 
A linear combination of two pure states, both of which are
entangled, viz. having a non-zero entropy for a subsystem, can
exhibit no entropy at all. And the reverse situation is also
possible, two pure states with no entanglement, can give rise to
entanglement on superposition. 

For a system consisting of a large number of qubits, how different
pairs share entanglement in a pure state cannot easily be specified,
even with known diagonal and off-diagonal correlation functions. 
We will show below, for the ground state Heisenberg antiferromagnet
with zero total spin, the concurrence measure can be
specified in terms of the diagonal correlation function alone.
In a pure state of many qubits, a subsystem of two qubits, in general,  
will be in a mixed state. A mixed state density matrix can be written as
a decomposition over pure state density matrices, with a large number of
possible decompositions over pure states. The entanglement of the pair
of qubits is the average entanglement of a decomposition, minimized over
all possible decompositions.  
Starting from a given many-qubit state, with a density matrix 
$\rho=|\psi><\psi|$, the reduced density matrix 
$R_{ij}$ for
a pair of qubits is constructed by performing a partial trace over the
rest of the qubits to be eliminated, $R_{ij}=tr \rho$. In general the
reduced density matrix represents a mixed state for the pair of sites
labeled $(i,j)$. The von Neumann entropy calculated from the eigenvalues
$r_n$
of $R_{ij}$, as $-\sum r_n \log_2 r_n$ quantifies entanglement of this
pair with the rest of the qubits. The concurrence measure\cite{Wootters}
has the important information as to how these qubits are entangled among
themselves, and is given as
\begin{equation}
C_{ij}=~{\rm max}~ (0, \lambda_1^{1/2}-\lambda_2^{1/2}-\lambda_3^{1/2}-
\lambda_4^{1/2}).
\end{equation} 
In the above $\lambda_i$ are the eigenvalues in decreasing order of the
matrix $R \hat R$, where $\hat R$ is the time-reversed matrix,
$\hat R=\sigma_y \times \sigma_y R^* \sigma_y \times \sigma_y$. 

The two-site concurrence depends only
on the structure of entanglement of a many-qubit state
without reference to a Hamiltonian. However, we would like to investigate
the pairwise entanglement in the ground state of the Heisenberg Hamiltonian. 
We will study the entanglement in the ground state Heisenberg antiferromagnet,
with a Hamiltonian
\begin{equation}
H= \sum_{<i,j>} {1\over2} (S_i^+ S_{j}^- +{S_i^- S_j^+}) + S_i^z
S_{j}^z,\end{equation}
The sum is over all
nearest-neighbor pairs on a given lattice. Working in a
diagonal basis of $S_i^z$ for every site $i$, there are two states
per site, $viz$., $\uparrow$, and $\downarrow$. A
many-spin state can be characterized by the number of down spins,
as the total z-component of the spin, $S^z=\sum S_i^z$, is a good quantum 
number, in addition to the
total spin quantum number $S$. The above Hamiltonian is also invariant
under time reversal. The ground state will belong to $S^z=0$ 
subspace, as this sector has a representation state from every spin
sector labeled by $S=0,1,..N/2$. 
For most lattices, and finite clusters the ground state belongs to
$S=0$, for example the linear chain, the square lattice, the triangular
lattice, Kagome' lattice, and finite clusters that will be considered below.
However, the ground state energy and the wave function are known only for
the case of the linear chain through the Bethe Anzatz\cite{Skryabin}.
The highest-energy state has $S=N/2$, which is the maximum value for the
total spin, with a degeneracy of $N+1$ corresponding
to the different values of $S^z=-N/2,..N/2$. This maximal-spin state would
be the ground state, if the interaction is ferromagnetic in the above
Hamiltonian (by changing the overall sign of the Hamiltonian).

\section{ Concurrences in $S=0$ states}

Let us consider a state with m down spins
\begin{equation} |\psi>=\sum \psi
(x_1..x_m)|x_1..x_m>\end{equation} where the ordered labels $x_1..x_m$ denote 
the locations
of the down spins. 
Let us rewrite the state as
\begin{equation}|\psi>=\sum_{s_1..s_N}\psi
(s_1..s_N)|s_1..s_N>\end{equation} where $s_i$ is the eigenvalue
of $S_i^z$ for the i'th qubit.  The reduced density matrix $R_{ij}$
of two sites $i$ and $j$ ($j>i$) has the matrix elements
\begin{equation}
R_{s_i,s_j} ^{s_i^{\prime},s_j^{\prime}}=\sum_{s_1..s_n}
\psi^{\star} (s_1,..s_i^{\prime}..s_j^{\prime}..s_n)\psi (s_1..s_i..s_j..s_n)
\end{equation}
where the sum is over all spin variables except at sites $i$ and $j.$ 
Since the many-qubit state has a definite eigenvalue for $S^z$, we
have $[R_{ij},S_i^z+S_j^z]=0$. This in turn implies the following structure
for the reduced density matrix
\begin{equation}R_{ij}=\left(\begin{array}{cccc}
        v_{ij}&0&0&0\\
        0& w_{1ij}&z_{ij}^{*}&0\\
        0&z_{ij}&w_{2ij}&0\\
        0&0&0&u_{ij} \end{array} \right).\end{equation}
In the above we used the two-qubit basis states $
|\uparrow\uparrow>,
|\uparrow\downarrow>,
|\downarrow\uparrow>,
|\downarrow\downarrow>,$ where $|\uparrow>,|\downarrow>$ stand for eigenstates
of $S_i^z$ with eigenvalues $1/2,-1/2$ respectively.

The matrix elements can be expressed in terms of the expectation values,
$<A>\equiv <\psi|A|\psi>$, as
\begin{equation}u_{i,j}=<({1\over2}-S^z_i)({1\over2}-S^z_j)>,v_{i,j}=
<({1\over2}+S^z_i)({1\over2}+S^z_j)>,\end{equation}
\begin{equation}z_{i,j}=<S^+_jS^-_i>,\end{equation}
\begin{equation}w_{1ij}=<({1\over2}+S^z_i)({1\over2}-S^z_j)>,
w_{2ij}=<({1\over2}-S^z_i)({1\over2}+S^z_j)>.\end{equation} 
The diagonal matrix elements are simply related to the diagonal
correlation functions, and the off-diagonal matrix element is
just the off-diagonal correlation function. The concurrence for the two
sites has now a simpler form
\begin{equation} C_{ij}=2~ {\rm max}~ (0, |z_{ij}|-\sqrt{u_{ij}v_{ij}}).
\end{equation}
As can easily be seen from above, the concurrence measure uses both
diagonal and off-diagonal correlation
functions. Whether or not two sites have a non-zero concurrence is
not at all intuitive, given a specific state with known correlation functions.
For a long-ranged concurrence, the necessary condition is the existence
of off-diagonal long range order (ODLRO). If there
is no off-diagonal long range order (ODLRO), $z_{ij}\rightarrow
0,$ as $|\vec r_i-\vec r_j|\rightarrow \infty$, the concurrence of
two sites far apart would go to zero. Thus, the existence of ODLRO is a
necessary condition for a long-ranged concurrence. However, even
with a ODLRO, long-ranged concurrence may be absent, if
$|z_{ij}|<\sqrt{u_{ij}v_{ij}}$. We will see below that the pairwise
concurrence is short ranged in the ground state of Heisenberg spin systems.
For a system with a large number of qubits, only the nearest-neighbor
concurrence seems to be nonzero.

Let us consider a nondegenerate $S=0$ state, which can be represented in
$S^z=0$ subspace. Because of the time-reversal symmetry of the Hamiltonian,
the wave function can be chosen to be real. This would imply that the
off-diagonal
matrix element $z_{ij}$ of the reduced density matrix will be real. Further,
the rotational symmetry of the Hamiltonian would imply, for the $S=0$
sector, $<S_i^xS_j^x>=<S_i^yS_j^y>=<S_i^zS_j^z>$.
Exploiting this property and denoting the diagonal correlation function by
$\Gamma_{ij}=<S_i^zS_j^z>$, we have
\begin{equation}
z_{ij}=<S_j^+S_i^->=2\Gamma_{ij}.
\end{equation}
The time-reversal invariance also implies, $<S_i^z>=0$, in this state. This
will simplify the diagonal matrix elements of $R_{ij}$.
Thus the diagonal elements of the reduced density matrix are
\begin{equation}
v_{ij}=u_{ij}={1\over4}+\Gamma_{ij}.
\end{equation} 
In addition the diagonal correlation function has the property $|\Gamma_{ij}|
\le 1/4$, since we are dealing with spin-1/2 species. Now the concurrence
can be specified completely in terms of the diagonal correlation function
alone.

The diagonal correlation function between two arbitrary spins can take both
positive or negative values. For a pair of qubits with $\Gamma_{ij}>0$,
$C_{ij}=2 {\rm max}(0,2\Gamma_{ij}-{1\over4}-\Gamma_{ij})=0.$
The above argument would imply  that the concurrence is zero for a pair of 
qubits, if their diagonal correlation function is positive, in a nondegenerate
$S=0$ state. On a bipartite lattice, if two spins belong to the same 
sublattice, then the diagonal correlation function would be positive,
implying the concurrence here is zero.
On the other hand, when $\Gamma_{ij} <0$, then we have $\sqrt{u_{ij} v_{ij}}=
1/4-|\Gamma_{ij}|$,
so that the concurrence is given by
$C_{ij}=2 {\rm max} (0,3|\Gamma_{ij}|-{1\over4})=6(|\Gamma_{ij}|-{1\over 12})$
if positive, and otherwise zero. This gives us a simple test for a nonzero
concurrence,  viz. if $\Gamma_{ij}<0,|\Gamma_{ij}|>1/12$. Thus in a 
nondegenerate $S=0$ state, the concurrence for a pair of spins is specified
entirely by their diagonal correlation function, as
\begin{eqnarray}
C_{ij} &&=0,~~~{\rm for~} \Gamma_{ij}>0 \nonumber \\
&&=0,~~~{\rm for~} \Gamma_{ij}<0,|\Gamma_{ij}|<{1\over 12} \nonumber \\
&&=6 (|\Gamma_{ij}|-{1\over12}),~~~{\rm for~} \Gamma_{ij}<0,|\Gamma_{ij}|
>{1\over 12} 
\end{eqnarray}
The above formula substantially simplifies the procedure of calculating
the pairwise concurrences in a many-qubit $S=0$ state. We simply calculate
the diagonal spin-spin correlation function, and the concurrence is completely
specified from the formula outlined. In the following section, we will
effectively use the formula to study the entanglement distribution and
sharing among various pairs in finite clusters and higher dimensional
lattices.
 
\begin{figure}
\includegraphics[angle=0,width=7cm]{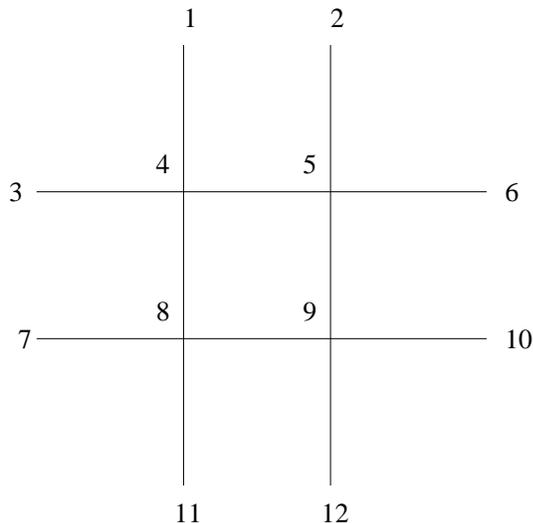}
\caption{A Tic-Tac-Toe cluster of twelve atoms. There are eight
inequivalent pairs here. Only two inequivalent
pairs of sites have nonzero concurrences in the ground state:
$C_{1,2}=0.08,C_{1,4}=0.41$.}
\end{figure}

Let us turn now to computing the actual values of the concurrences in
one-dimensional systems.
Now, for a closed chain of 12 sites, the above implies, denoting $C_{ij}=C_r$
where $r=j-i$, in the ground state 
$C_1=0.398,$ and $ C_r=0$ for $r>1$,
where we have used the numerically computed diagonal correlation
functions, $\Gamma_1=-0.1496,\Gamma_2=0.0626,\Gamma_3=-0.0553~ etc.$. As the 
size of the chain is increased,
the magnitude of the correlation function $\Gamma_r$ will take the limiting 
value from above. This would imply $C_r=0$, for $r>1$ for an infinite chain.  
$C_1$ can be estimated from the
ground state energy in the thermodynamic limit. The ground state
energy of the Heisenberg Hamiltonian in general can be written as
\begin{equation}
E_g\equiv e_gN=3N_n\Gamma_1
\end{equation}
where $e_g$ is the ground state energy per site, and $N_n$  
the number of nearest-neighbor pairs of spins. 
Thus the nearest-neighbor concurrence can be written as
\begin{equation}
C_1= 6 (|e_g| {N\over 3 N_n}-{1\over12})
\end{equation}
when positive, and otherwise zero. For a linear-chain antiferromagnet ground
state, a similar formula has been derived\cite{Connor}. 
The ground state
energy is known from Bethe Ansatz, $e_g=\ln 2 -1/4$, and $N_n=N$, which
determines the nearest-neighbor concurrence, $C_1=2\ln 2-1\approx 0.386.$ 

\section{Finite clusters and extended two-dimensional lattices}

In two-dimensional systems there are no analytical results for the ground
state of Heisenberg antiferromagnet systems, like the Bethe-ansatz solution
for the linear chain. Here, we will take recourse to numerical diagonalizations 
of finite clusters, and use variational estimates for the ground state
energies on various extended lattices.
The main difference between two-dimensional systems and the linear
chain considered above is the number of nearest neighbors, apart from
various local lattice structural differences. To gain insight into the
entanglement distribution in two dimensions, we will numerically study
below various finite clusters, which are drawn from square, 
triangular and Kagome' lattices. 

Let us first consider a toy problem of a Tic-Tac-Toe cluster of twelve sites, 
as shown in Fig.1. This cluster is drawn from a square lattice. Here there are
two inquivalent sites, site 1 is equivalent to
the other sites on the boundary, and site 4 is equivalent to the other
sites forming the vertices of the inside square.  There are eight
inequivalent pairs of sites: (1,2), (1,3),(1,4),(1,5),(1,9),(1,10),(4,5),
(4,9). From the numerical exact diagonalization for the $S=0$ ground
state, only two inequivalent pairs have nonzero concurrence:
$C_{1,2}=0.08,C_{1,4}=0.41$, and $C_{1,3}=C_{1,5}=C_{1,9},C_{1,10}=C_{4,5}=
C_{4,9}=0.$ In this finite cluster, the third-neighbor bonds, (1,2),(1,6),
(1,7),(1,11) have concurrence equal to 0.08, where as the second-neighbor
bonds (1,5),(1,8),(1,3) etc. have zero concurrence. This can be ascribed
to finite-size effects, and to the fact that this cluster does not have
the same symmetry as the square lattice.
 
Let us now consider a 16-site square lattice cluster, with periodic boundary 
conditions. Here all sites are equivalent, with coordination number equal
to four as on an infinite square lattice. This cluster has also the
bipartite structure of the square lattice.
There are only four inequivalent pairs, due to the periodic 
boundary conditions. For a given site, the second and third neighbors sit
on the same sublattice, implying that the diagonal correlation function is
positive, $\Gamma=0.071$ from the numerical calculation, and hence a 
zero concurrence. 
The numerical values of the correlation function are  -0.117 and -0.067 for
the nearest and furthest neighbor pairs respectively. 
Only the nearest-neighbor concurrence is
nonzero, $C_1=0.202$ which is almost half the value we got for the linear chain.
To study larger clusters, one has to increase the number of sites to 24, to
keep the bipartite structure intact. Then exact numerical diagonalization is
quite difficult for such large clusters.

\begin{figure*}
\includegraphics[angle=0,width=8cm,height=6cm]{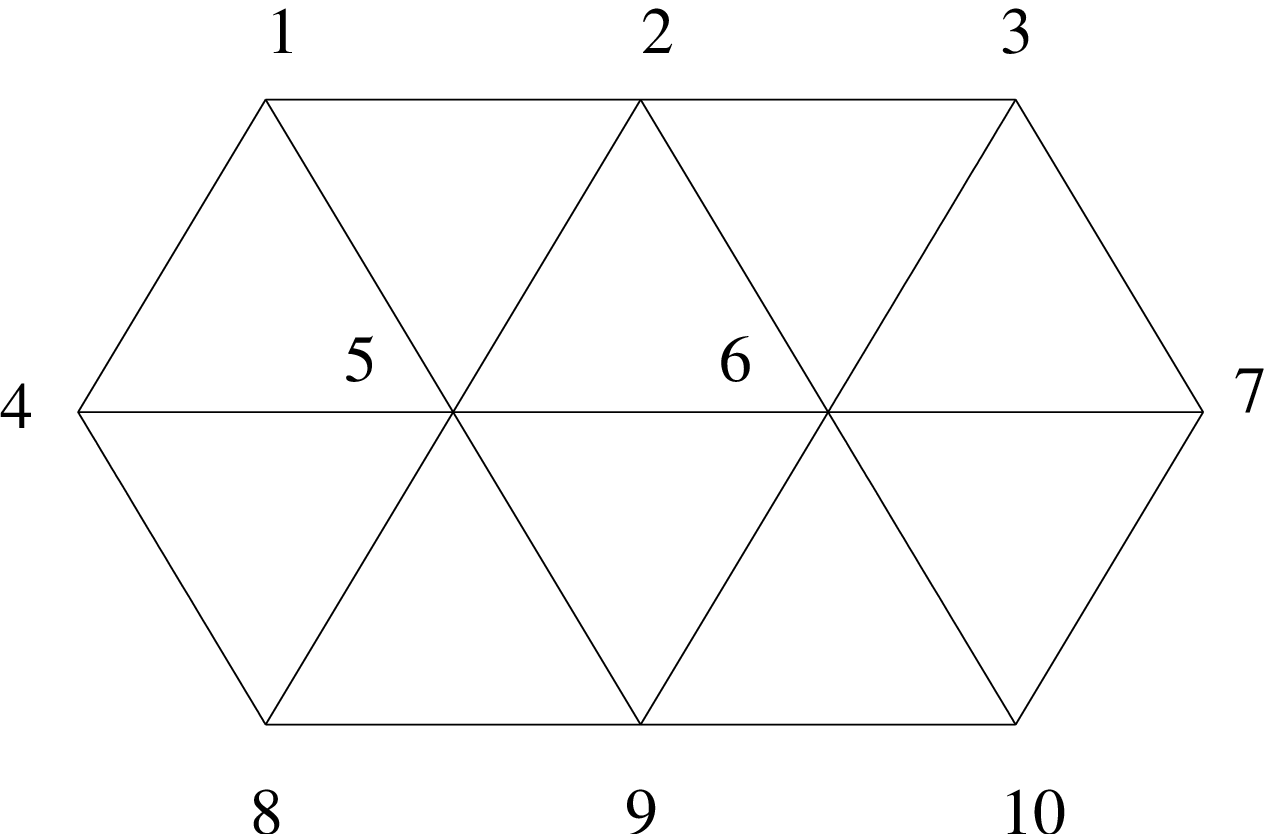}
\includegraphics[angle=0,width=8cm,height=7.5cm]{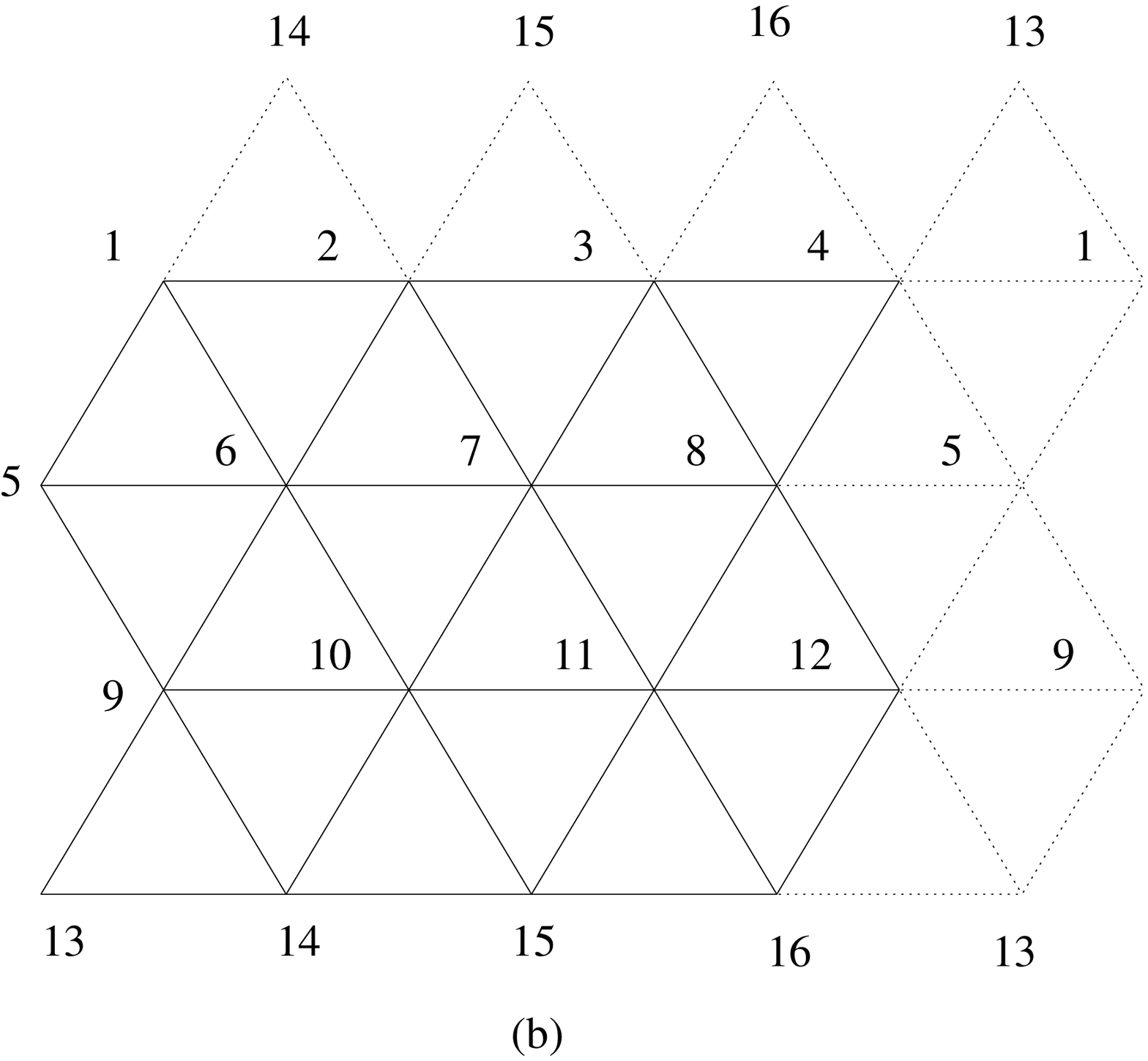} 
\caption{(a) A triangular-lattice cluster with ten atoms. Here all the
sites are not equivalent. There are four inequivalent sites labeled
1, 2, 4, 5.
(b) A 16-site triangular cluster, with periodic boundary
conditions. The extra nearest-neighbor bonds arising from the boundary
conditions are shown with dotted lines. Here all sites are equivalent, with
six nearest neighbors, like in the infinite triangular lattice. However,
the horizontal bonds (sixteen in number) are not equivalent to the slanted
bonds (32 in number).}
\end{figure*}
For the infinite square lattice the exact ground state energy is not known 
from analytical
solutions, however, there are excellent estimates of upper bounds on $e_g$. For
the square-lattice ground state\cite{Huse}, we have an upper bound
estimate $|e_g|\approx 0.66$. 
This in turn, along with $N_n=2N$,  implies
\begin{equation}
C_1= |e_g|-{1\over 2}\approx 0.16 ~~~{\rm for~square~~lattice}.
\end{equation} 
And the next-neighbor concurrence is zero, as the two spins sit on one 
sublattice of the square lattice. We conjecture that $C_r=0, r>1$ for square 
lattice also. This is because the concurrence reduces when the number of
neighbors 
increases, as is evident from the nearest-neighbor concurrence (it reduced from
0.386 to about 0.16, as the number of nearest neighbors increased from 2 to 4,
from the linear chain to the square lattice). And in conjunction, we have
the fourth-neighbor concurrence zero from the sixteen-site cluster studied
above.

Let us now turn our attention to the triangular lattice, which has
an increased number of nearest neighbors, along with a new ingredient of
frustration among the spins on a triangular block, for minimization of
energy. Again we take recourse to numerical diagonalization of finite
clusters, in the absence of exact results for the ground state Heisenberg
antiferromagnet.
We will consider two different clusters here. Let us first
look at a ten-site cluster shown in Fig.2a, which is drawn from a
triangular lattice. Though all sites are not equivalent here, but
it has the
basic triangular and hexagonal block structure of a triangular lattice.
From the numerical exact diagonalization for the ground state, only
three inequivalent pairs have nonzero concurrence. The horizontal 
nearest-neighbor bond on the boundary has a concurrence, $C_{1,2}=0.26,$ 
along with the three other equivalent bonds. The slanted nearest-neighbor
bonds on the boundary have a concurrence, $C_{1,4}=0.347.$ The bond at
the center has a concurrence, $C_{5,6}=0.336.$ All the other
nearest-neighbor bonds have zero concurrence, along with all other pair
concurrences between sites further apart. The magnitude of the 
nearest-neighbor pair concurrence
has reduced here from the linear chain on account of the increased number
of neighbors and the frustration associated with the triangular blocks. 

Let us now consider a 16-site triangular cluster with periodic boundary
conditions, shown in Fig.2b.
The sites on the boundary get
extra neighbors, due to the boundary conditions, indicated by dotted lines
in the figure. For instance,
site labeled as 1 has extra neighbors in sites labeled as 4,13 and 14, 
and so on for the other sites on the boundary (see figure). Now all the
sites are equivalent, with six nearest neighbors as on an extended 
triangular lattice. This cluster has the ingredients of an infinite
triangular lattice: six coordination, triangular block structure and
associated frustration, equivalence of all the sites.
One important ingredient that is missing here is the
tripartite structure. There are two inequivalent nearest-neighbor bonds
here, horizontal and slanted bonds. 
Starting from any site and traversing four horizontal bonds, one can
return to the original site, whereas by traversing four slanted bonds
one cannot return, as can be seen from the figure. From the numerical
diagonalization, we get for the concurrence for the 16 horizontal bonds,
$C_{1,2}=0,$ and for the 32 slanted bonds, $C_{1,5}=0.05$. The
concurrence is zero for all other pairs of sites further apart. Here, the
concurrence is furthur reduced from that of the square-lattice clusters
due to increased neighbors and frustration. This discrepancy between
the horizontal bonds and the slanted bonds will disappear as the number of
sites is increased, and the tripartite structure restored for an extended
triangular lattice. 

For an extended triangular lattice,
we will use the variational estimates for the ground state energy, for
calculation of nearest-neighbor concurrences.
For a triangular lattice, we have $N_n=3N$, which implies
\begin{equation}
C_1=2({|e_g|\over 3}-{1\over 4}) ~~{\rm for ~triangular~~lattice}
\end{equation} 
if positive, and otherwise zero. The upper bound estimate for the ground state 
energy\cite{Huse,Subrahmanyam} is $|e_g|\approx 0.53$. Hence, the 
nearest-neighbor 
concurrence for the triangular lattice is zero. The next-neighbor concurrence 
cannot be argued to be zero based on a bipartite
structure, as we did above for the linear chain and the square lattice. 
However, we can argue that the absolute value of the correlation function would 
decrease as the separation increase. Also, from the 16-site cluster with
six coordination discussed above, other than nearest-neighbor concurrences
are vanishing. We expect as the size is increased, they will still be zero.
This would imply $C_r=0$ for all separations on a triangular lattice.  

We would like to study the Kagome' lattice clusters now. The obvious cluster
that can be considered is a David star cluster of twelve sites shown in Fig.3,
drawn from a Kagome' net. This cluster too has triangular blocks, along with
the associated frustration, apart from unfilled hexagonal blocks unlike the
triangular lattice. Numerically it is seen that this cluster has two degenerate $S=0$ ground states. The pair concurrences in one of the ground states, for
the nearest-neighbor bonds can be calculated from the reduced density
matrices. The only nonzero nearest-neighbor concurreces are for alternate
bonds on the boundary, traversing clockwise, we have $C_{14}=C_{5,7}=C_{11,10}=
C_{12,9}=C_{8,6}=C_{2,3}=0.55.$ Similarly the other ground state has nonzero
concurrences only for pairs (1,3), (2,6), (8,9), (12,10), (11,7),(5,4).
This is because these ground states have a large amplitude for singlets for
these pairs. This can be seen easily by viewing the cluster shown in figure
as a closed chain with every alternate site having a next-neighbor interaction.
If every site has a next-neighbor interaction whose strength is half that
of the nearest-neighbor interaction, then this reduces to Majumdar-Ghosh
model, which has a two-fold degenerate valence-bond ground state\cite{Majumdar}.
But here, the next-neighbor interaction strength has been transfered to every
second next-neighbor bond. But, still, the amplitude of the valence-bond
configuration is predominant, giving rise to dominant contribution for
concurrences only for alternate bonds.
All the other pairs have zero concurrence. However, since the
ground state is degenerate, one can change these concurrences by suitable
linear combinations of the degenerate states.
Larger clusters drawn from the Kagome' lattice, keeping the ingredients intact,
are too large for exact diagonalization. Again, we will turn to the 
variational estimates for the ground state energy, to infer the
nearest-neighbor concurrences.

\begin{figure}
\includegraphics[angle=0,width=7cm,height=7.5cm]{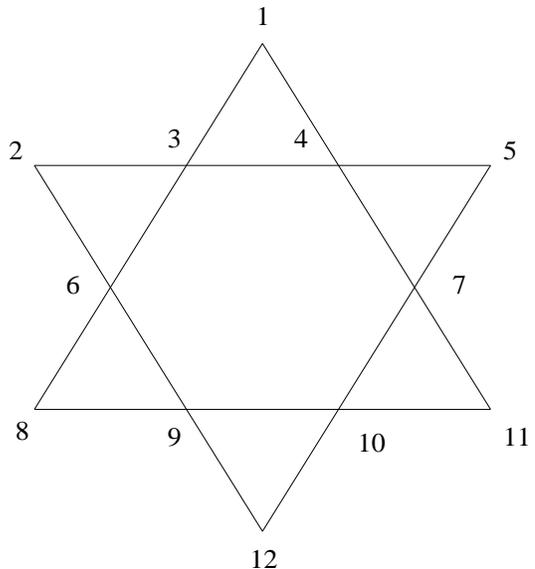}
\caption{A Kagome'-lattice David star cluster with twelve atoms. The ground
state here is two-fold degenerate, with large amplitude for valance-bond
configurations.}
\end{figure}
For the case of an extended Kagome' lattice, we have 
$N_n=2N$, and the estimate for the ground state energy\cite{Elser,Subrahmanyam},
$|e_g|\approx 5/12$, would imply that the nearest-neighbor
concurrence is zero. Analogous to the triangular lattice, we can argue $C_r=0$
for all separations. Though the square 
lattice and the Kagome' lattice have the same number of nearest neighbors, 
$C_1$ is zero for Kagome lattice. This we can attribute to the frustration
present in the Kagome' case, due to the presence of triangular 
blocks\cite{Subrahmanyam}. 

We can increase the number of nearest neighbors by considering a 
$d$-dimensional hypercubic lattice, with $2d$ nearest neighbors for every site.
For large $d$, the ground state energy has an upper bound $|e_g|\approx d/4$, 
and
$N_n=Nd$ which will yield $C=0$ for the nearest neighbors. Similarly we expect
all pairs of sites to have zero concurrence. This is consistent with our
argument, that an increase in the number of neighbors will result in a
reduction in the nearest-neighbor concurrences. We can also increase the
effective number of
neighbors by including a long-ranged interaction, by making the sum in the
Heisenberg Hamiltonian (Eq.2) run over all pairs. Then each qubit interacts
with all
other qubits by the same strength. This gives us $H=\sum \vec S_i\cdot\vec S_j
=(S^2-3N/4)/2$, the eigenvalue depending only on the total spin quantum number
$S$. A larger value for the total spin will lead to a larger energy.
The lowest eigenvalue is  for $S=0,$ $e_g=-3/8$. Now, we
also have $e_gN=3N(N-1)\Gamma_{ij}/2$, as all the two-point correlators
would be identical, and $\vec S_i\cdot \vec S_j=3 \Gamma_{ij}$ for a $S=0$
state.  This yields $C=0$, for all pairs of qubits.  

\section{Maximal-spin states}
Let us turn our attention to the maximal-spin states, with $S=N/2$, which are 
the ground 
states for a ferromagnetic interaction, and are maximal-energy states for the
antiferromagnet, with an energy $E_{max}=N/4.$ The structure of the
maximal-spin states will be same for all clusters and lattices, irrespective
of coordination number and local structure. 
There are $N+1$ such states,
with one state each in a subspace with a given value $S^z=-N/2,..N/2$. 
Let us write $S^z={N/2-m}$
where $m$ is the number of down spins in each basis state, $m=0,1..N$ for 
various subspaces. It is straightforward to write down the eigenstate for
any lattice as
\begin{equation}
|S={N\over2},S^z={N\over 2}-m>= {1\over ^NC_m}\sum |i>
\end{equation}
where the sum is over $^NC_m$ basis states each with $m$ down spins exactly.
Since every basis state is given the same weight, the reduced density matrix
for any pair of spins will have the same matrix elements, and the correlation
functions, both the diagonal and the off-diagonal, will be same for every
pair. This implies $C_{ij}=C$ for all the $N_p=N(N-1)/2$ pairs. The diagonal 
correlation 
function, thus, can be written as $\Gamma_{ij}=(1/N_p)\sum_{\rm all pairs}
<S_i^zS_j^z>=
(S{^z}^2-N/4)/N(N-1)$. 
For values of $S^z\sim O(N)$, 
there is a long-ranged order. This would
cause a decrease in the concurrence for a given pair, with a concurrence
$\sim O(1/N)$. However, the
average concurrence for this case is more, as we will see below, than the 
concurrence in the
$S=0$ state, that we discussed for various lattice systems above. 
Thus, the off-diagonal matrix elements of $R_{ij}$ are evaluated as
\begin{eqnarray}
u_{ij}&&= {({N\over2} -S^z)({N\over2}-S^z-1)\over N(N-1)}\nonumber \\
v_{ij}&&
={({N\over2} +S^z)({N\over2}+S^z-1)\over N(N-1)}.\nonumber \\
\end{eqnarray}

Now, for the off-diagonal matrix element $z_{ij}=<S_j^+S_i^->$, the 
contributions
come from all configurations with $\downarrow$ at site $j$ and $\uparrow$
at site $i$. There are $N-2C_{m-2}$ such configurations. Hence,
\begin{equation}
z_{ij}
={({N\over2} -S^z)({N\over2}+S^z)\over N(N-1)}.
\end{equation}
Putting all things together, the concurrence for any pair of qubits is,
with $S=N/2,S^z=N/2-m$,
\begin{equation}
C(m)={2m(N-m)\over N(N-1)}
\{ 1-\sqrt{(m-1)(N-1-m)
\over m(N-m)}\}.
\end{equation}
A similar formula has been given in\cite{Wang}.
Some values of the function are $C(S=N/2,S^z=\pm N/2)=0$, and
\begin{eqnarray}
C(S={N\over2},S^z=0)={1\over N-1} \nonumber \\
C(S={N\over2},|S^z|={N\over 2}-1)={2\over N}.
\end{eqnarray}

The concurrence is maximum when the number of down spins is either 1 or N-1.
Though all pairs have the same concurrence, however, the concurrence is
not long ranged, since in the thermodynamic limit, $N\rightarrow\infty$, the
concurrence goes to zero as $1/N$.
But, in comparison the state with $S=N/2,S^z=0$ has a larger average 
concurrence than the state with $S=0=S^z$ that we studied earlier, where
the best average concurrence for the linear chain is $<C>\approx 0.8/(N-1)$.
Thus, increasing the spin from $S=0$, for the ground state, to $S=N/2$, for
the maximal-energy state of the antiferromagnet, the nearest-neighbor
concurrence has decreased from $O(1)$ to $O(1/N)$. At the same time, the
average over all pairs has improved, implying a better entanglement sharing
in the maximal-spin states.
Thus, the  maximal-spin states with $S=N/2,|S^z|<N/2$ have a better 
entanglement sharing
among the qubits, with the best sharing in the states $S=N/2,|S^z|=N/2-1$,
with an average concurrence $<C>=2/N$. The sectors with $S=N/2-1, N/2-2$,
namely the one-magnon and two-magnon states, have been 
investigated\cite{Arul, Subrah}.
%and it has been seen that $<C>$ is slightly greater than $2/N$ in a two-magnon
%bound state.  
However, it remains to see if states with intermediate values of energy
and/or with intermediate values of spin $0<S<N/2-2$,
can exhibit more average concurrence. Most important states are the first
excited sates of the antiferromagnet, with $S=1$, where the concurrence
results are yet to be known. It would be interesting to investigate
the concurrence as a function of $S, S^z$ in all spin sectors. 

\section{Conclusion}
We have investigated the quantum entanglement sharing in Heisenberg
antiferromagnets using the concurrence measure. For S=0 unique ground states, 
a simple formula relates the concurrence between an arbitrary pair of spins
to their diagonal correlation function. This substantially simplifies the
calculation of concurrences, both numerically and analytically.
The nearest-neighbor concurrence is directly related to the ground state
energy. For larger number of nearest-neighbors, the concurrence is smaller,
as has been seen from studying a number of finite clusters, and from the
study of large-d hypercubic lattice and longer-ranged interactions.
It has been argued that other than nearest-neighbor pairs have zero
concurrence in the ground state for the linear chain and the square lattice.
It is shown that the nearest-neighbor concurrences also vanish in the ground
state of triangular and Kagome lattice antiferromagnets.
For maximal-spin states, explicit formulas are given for the pairwise 
concurrences, and the states with
$S=N/2, S^z=N/2-1$ show maximum entanglement sharing.

It is a great pleasure to thank Professors Arul Lakshminarayan and
V. Ravishankar for extended discussions.

\end{document}